\documentclass[12pt]{iopart}

\usepackage{iopams}
\usepackage{setstack}

\usepackage{graphicx}
\usepackage{float}
\usepackage{hyperref}

\begin{document}

\title[Systematic construction of density functionals based on MPS computations]{Systematic construction of density functionals based on matrix product state computations}

\author{Michael Lubasch$^{1}$\footnote{Present address: Department of Physics, University of Oxford, Parks Road, Oxford OX1 3PU, UK.}, Johanna I Fuks$^{2}$, Heiko Appel$^{3, 4}$, Angel Rubio$^{3, 4}$, J Ignacio Cirac$^{1}$ and Mari-Carmen Ba\~{n}uls$^{1}$}
\address{$^{1}$ Max-Planck-Institut f\"{u}r Quantenoptik, Hans-Kopfermann-Stra\ss{}e 1, 85748 Garching, Germany}
\address{$^{2}$ Department of Physics and Astronomy, Hunter College and the Graduate Center of the City University of New York, 695 Park Avenue, New York, New York 10065, USA}
\address{$^{3}$ Fritz-Haber-Institut der Max-Planck-Gesellschaft, Faradayweg 4-6, 14195 Berlin-Dahlem, Germany}
\address{$^{4}$ Max-Planck-Institut f\"{u}r Struktur und Dynamik der Materie, Luruper Chaussee 149, 22761 Hamburg, Germany}
\ead{michael.lubasch@mpq.mpg.de}

\begin{abstract}
We propose a systematic procedure for the approximation of density functionals in density functional theory that consists of two parts.
First, for the efficient approximation of a general density functional, we introduce an efficient ansatz whose non-locality can be increased systematically.
Second, we present a fitting strategy that is based on systematically increasing a reasonably chosen set of training densities.
We investigate our procedure in the context of strongly correlated fermions on a one-dimensional lattice in which we compute accurate training densities with the help of matrix product states.
Focusing on the exchange-correlation energy, we demonstrate how an efficient approximation can be found that includes and systematically improves beyond the local density approximation.
Importantly, this systematic improvement is shown for target densities that are quite different from the training densities.
\end{abstract}



\maketitle

\section{Introduction}

The formulation of quantum mechanics in terms of density functionals instead of wave functions, following the ground-breaking works of Hohenberg, Kohn, and Sham~\cite{Hohenberg, Kohn}, made numerical simulations of quantum mechanical systems ranging from the microscopic to the macroscopic world feasible~\cite{Dreizler, Kohn2, Jones}.
The usefulness of density functional theory (DFT) is certified by the number of works based on the original publications~\cite{Hohenberg, Kohn} and on later improvements of the exchange-correlation (xc) energy density functional $E_{\mathrm{xc}}$~\cite{Perdew, Lee, Becke, Perdew2, Becke2, Perdew3, Zhao}.

DFT in its most widely used form, namely Kohn-Sham (KS) DFT~\cite{Kohn}, requires the xc density functional in order to be able to compute ground state energies and densities.
By virtue of the Hohenberg-Kohn theorem~\cite{Hohenberg}, all ground state observables are functionals of the ground state density $n$, and so $E_{\mathrm{xc}}=E_{\mathrm{xc}}[n]$.
The ground state energy $E=E[n]$ of a system can be decomposed into a kinetic, an interaction and a potential part.
By means of a fictitious non-interacting system, namely the KS system, the non-interacting part of the kinetic energy, $T^{\mathrm{s}}=T^{\mathrm{s}}[n]$, can be obtained efficiently, which represents a large contribution to the full interacting kinetic energy $T$.
Further, part of the interaction energy is accounted for by the Hartree energy $E_{\mathrm{H}}[n]$.
The potential part $E_{V}[n]$ can be exactly computed efficiently for any ground state density $n$.
Finally, the remaining part of the total ground state energy defines the xc density functional, $E_{\mathrm{xc}}[n]:=E[n]-T^{\mathrm{s}}[n]-E_{\mathrm{H}}[n]-E_{\mathrm{V}}[n]$.
DFT is in principle exact, but in practice determining the precise form of the xc density functional is QMA-hard~\cite{Schuch}.
Therefore, KS DFT can only make use of approximations of $E_{\mathrm{xc}}$.
The enormous success of DFT is thus deeply connected to the successful construction of good approximations for the xc energy density functional.

In the history of DFT and quest for a universally applicable approximate $E_{\mathrm{xc}}$~\cite{Peverati}, mainly two different paths have been followed: one is the non-empirical approach pioneered by Perdew~\cite{Perdew4} and the other is the semi-empirical approach initiated by Becke~\cite{Becke}.
The non-empirical approach makes use of exact conditions, that a physical system must fulfill, to find approximations for the xc density functional.
Within this approach a ``Jacob's ladder'' of functionals was built where each functional on a higher rung of the ladder is supposed to improve upon the ones on the lower rungs~\cite{Kurth, Perdew5}.
On the lowest rung of the ``Jacob's ladder'' resides the local density approximation (LDA), which was already introduced by Kohn and Sham in~\cite{Kohn}.
The higher rungs are supposed to systematically improve upon the LDA, which, in practice, does not always happen~\cite{Perdew5}.
Additionally, at the moment, the more precise functionals on the higher rungs are so much more difficult to compute that further improvements of DFT following this non-empirical approach seem very hard to achieve.
In the semi-empirical approach, an ansatz for the functional form of $E_{\mathrm{xc}}$ is fitted using experimental data, accurate theoretical reference data, or other constraints.
However, often relatively small training sets are used in these fits and then the resulting functionals can be biased towards their training~\cite{Peverati}.

Alternatively, we might obtain further improvements of $E_{\mathrm{xc}}$ away from but using concepts of both the semi-empirical and the non-empirical approach, e.g.\ by using a large set of accurate training densities and corresponding values of $E_{\mathrm{xc}}$, and by fitting an efficient ansatz to these data that includes some exact conditions.
Obviously, a difficulty of this alternative scheme is that it requires a possibly large number of accurate solutions for the quantum many-body problem.
However, nowadays, tensor network states provide precise results for quantum many-body systems, e.g.\ \cite{Verstraete, Orus}, in particular with respect to ground state properties.
We remark that tensor network methods are currently limited to low-dimensional, i.e.\ one- and some two-dimensional, quantum lattice problems while DFT usually handles three-dimensional continuous quantum systems.
Since DFT can be applied to a wide range of realistic quantum systems, it is a useful algorithm for a large community and thus worth improving.

In this article, we want to analyze the feasibility of constructing an approximate xc density functional of a specific form, when large training sets of ground state densities and corresponding values of $E_{\mathrm{xc}}$ are available.
The specific form for the ansatz of our approximation is inspired by the non-empirical approach~\cite{Perdew5}: it includes the LDA~\cite{Kohn, Giuliani} and allows a systematic improvement beyond it.
For this feasibility study, we focus on discrete lattice problems and the one-dimensional case, and we use matrix product states (MPS) for the computation of accurate ground state energies and densities~\cite{White, Schollwoeck}.
The specific discrete lattice problem considered here can be derived from discretization of continuous space, i.e.\ the usual scenario of DFT.
Then our ansatz can be seen as the discretized version of a continuous function.
Although we could approach the continuum solution by successively decreasing the discretization, taking the continuum limit is beyond the scope of this work.

The structure of this article is as follows.
In \sref{sec:model} we introduce the considered Hamiltonian and observables.
The corresponding exact LDA is presented in \sref{sec:exactLDA}.
We then propose, fit, and assess our ansatz in \sref{sec:ouransatz}.
Finally, in \sref{sec:conclusions} we conclude this work and give an outlook.

\section{Model}
\label{sec:model}

In the following, we consider two species of fermions with long-ranged soft-Coulomb interaction on a finite one-dimensional lattice of length $L$ with hard-wall boundary conditions, as represented by the Hamiltonian:
\begin{eqnarray}\label{eq:ham}
 \hat{H} := \hat{T} + \hat{W} + \hat{V}
\end{eqnarray}
with
\numparts\begin{eqnarray}\label{eq:hamT}
 \hat{T} := -t\sum_{l=1}^{L-1}\sum_{\sigma=\uparrow,\downarrow}(c_{l,\sigma}^{\dag}c_{l+1,\sigma}+c_{l+1,\sigma}^{\dag}c_{l,\sigma})\\\label{eq:hamW}
 \hat{W} := U\sum_{l=1}^{L}\Big(\hat{n}_{l,\uparrow}\hat{n}_{l,\downarrow}+\sum_{m=l+1}^{L} \frac{\hat{n}_{l}\hat{n}_{m}}{\sqrt{(m-l)^{2}+1}}\Big)\\
 \hat{V} := \sum_{l=1}^{L}(v_{l}^{\mathrm{ext}}-\mu)\hat{n}_{l} \qquad .
\end{eqnarray}\endnumparts
Here, $c_{l,\sigma}^{\dag}$ creates and $c_{l,\sigma}$ annihilates a fermion of species $\sigma=\,\,\uparrow,\,\downarrow$ on lattice site $l$, $\hat{n}_{l,\sigma}:=c_{l,\sigma}^{\dag}c_{l,\sigma}$ is the corresponding occupation number operator and $\hat{n}_{l}:=\hat{n}_{l,\uparrow}+\hat{n}_{l,\downarrow}$.
The total particle number is denoted by $N:=\langle\sum_{l=1}^{L}\hat{n}_{l}\rangle$.
We obtain ground states with different total particle number by choosing different values for the chemical potential $\mu$, which plays the role of a Lagrange multiplier fixing $N$.
Such a Hamiltonian can also describe the discretized continuous problem with lattice spacing $\Delta$ when in~\eref{eq:hamT} $t$ is replaced by $1/(2\Delta^{2})$ and in~\eref{eq:hamW} the denominator $\sqrt{(m-l)^{2}+1}$ is replaced by $\sqrt{(m-l)^{2}\Delta^{2}+1}$.
The solution for different discretizations can be very precisely computed with MPS~\cite{Stoudenmire, Wagner} and so our approach should yield highly accurate training densities.
If we would like to obtain the solution for continuous space, we would have to run our computations repeatedly with decreasing lattice spacing $\Delta$ and extrapolate our results to $\Delta=0$.
Here we see~\eref{eq:ham} as the Hamiltonian of the problem, and not a discrete version of a more fundamental one, thus we set $t=1/2$ and $U=1$ and fix the number of lattice sites to $L=21$ from now on.

On a finite lattice, densities $n_{l}:=\langle\hat{n}_{l}\rangle=\langle\hat{n}_{l,\uparrow}+\hat{n}_{l,\downarrow}\rangle$ can be written into a vector $\bi{n}:=(n_{1}, n_{2}, \ldots, n_{L})^{\mathrm{T}}$ - where $\mathrm{T}$ denotes the transpose - such that every density functional $F$ can be written as a function of such density vectors $F=F(\bi{n})$.
In the following, we will consider the \emph{universal Hohenberg-Kohn functional} $F_{\mathrm{HK}}(\bi{n})$, the \emph{Hartree-energy} $E_{\mathrm{H}}(\bi{n})$, and the \emph{non-interacting kinetic energy} $T^{\mathrm{s}}(\bi{n})$, e.g.\ \cite{Giuliani}.
For the above Hamiltonian~\eref{eq:ham} these functionals read:
\numparts\begin{eqnarray}\label{eq:funcFHK}
 F_{\mathrm{HK}}(\bi{n}) := E(\bi{n}) - \sum_{l=1}^{L}(v_{l}^{\mathrm{ext}}-\mu)n_{l}\\\label{eq:funcEH}
 E_{\mathrm{H}}(\bi{n}) := U\sum_{l=1}^{L}\Big(\frac{n_{l}^{2}}{4} + \sum_{m=l+1}^{L} \frac{n_{l}n_{m}}{\sqrt{(m-l)^{2}+1}}\Big)\\\label{eq:funcTS}
 T^{\mathrm{s}}(\bi{n}) := E^{\mathrm{s}}(\bi{n}) - \sum_{l=1}^{L}(v_{l}^{\mathrm{s}}-\mu^{\mathrm{s}})n_{l} \qquad ,
\end{eqnarray}\endnumparts
where $E(\bi{n})$ denotes the ground state energy of an interacting density $\bi{n}$, i.e.\ corresponding to~\eref{eq:ham} with $\hat{W}$, and $E^{\mathrm{s}}(\bi{n})$ denotes the ground state energy of a non-interacting density $\bi{n}$, i.e.\ corresponding to~\eref{eq:ham} without $\hat{W}$.
Knowing the values of these functionals~\eref{eq:funcFHK},~\eref{eq:funcEH}, and~\eref{eq:funcTS} for a particular density $\bi{n}$ allows to calculate the \emph{xc energy} $E_{\mathrm{xc}}$ for that density:
\begin{eqnarray}\label{eq:Exc}
 E_{\mathrm{xc}}(\bi{n}) := F_{\mathrm{HK}}(\bi{n})-E_{\mathrm{H}}(\bi{n})-T^{\mathrm{s}}(\bi{n}) \qquad .
\end{eqnarray}

However, given an arbitrary density vector $\bi{n}$, only $E_{\mathrm{H}}(\bi{n})$ is trivial to compute: $F_{\mathrm{HK}}(\bi{n})$ requires the knowledge of the external potential as a function of the density, $v_{l}^{\mathrm{ext}}=v_{l}^{\mathrm{ext}}(\bi{n})$, and $T^{\mathrm{s}}(\bi{n})$ requires the knowledge of the effective non-interacting Kohn-Sham potential $v_{l}^{\mathrm{s}}=v_{l}^{\mathrm{s}}(\bi{n})$.
This process of calculating the external potential $v_{l}(\bi{n})$, in which the ground state has the given density $\bi{n}$, is called \emph{inversion} and can be performed efficiently only in the non-interacting case or for two fermions~\cite{Peirs}.
In general, there exists no efficient inversion procedure for the interacting case and we use a slight modification of the iteration proposed in~\cite{Thiele, Stoudenmire2}:
Aiming at the target density $\bi{n}^{\mathrm{tar}}$, we iterate $v_{l}(i+1)=v_{l}(i)+\gamma(i)(n_{l}(i)-n_{l}^{\mathrm{tar}})$ until $||\bi{n}(i)-\bi{n}^{\mathrm{tar}}||$ - where $||\ldots||$ denotes Euclidean norm - is below a desired precision threshold.
Here, $\bi{n}(i)$ is the ground state density in the external potential $\bi{v}(i)$ at the iteration step $i$, and $\gamma(i)>0$ is adjusted during the iterations to speed up the convergence.
Since interacting inversion necessitates several ground state computations to attain an approximate solution, it is not efficient.
Even more sophisticated iteration schemes cannot circumvent that some densities require incredibly many iterations, i.e.\ ground state computations, until convergence~\cite{Wagner2}.
Therefore, in general, interacting inversion represents a computationally demanding task.

\section{Exact LDA}
\label{sec:exactLDA}

As the exact form of $E_{\mathrm{xc}}$ from~\eref{eq:Exc} is not known, in practice, approximations are used.
One of the simplest and most successful approximations is the LDA~\cite{Kohn, Giuliani}.

The exact LDA $e_{\mathrm{xc}}^{\mathrm{LDA}}$ is defined via the homogeneous electron gas, i.e.\ via exactly homogeneous densities $\bi{n}=(n_{1}, n_{2}, \ldots, n_{L})^{\mathrm{T}}=(n, n, \ldots, n)^{\mathrm{T}}$ in the thermodynamic limit $L \to \infty$~\cite{Kohn, Giuliani}:
\begin{eqnarray}\label{eq:excLDA}
 e_{\mathrm{xc}}^{\mathrm{LDA}}(n) := \lim_{L \to \infty} E_{\mathrm{xc}}(n, n, \ldots, n)/L \qquad .
\end{eqnarray}
This quantity is then used to approximate the xc energy of a finite system by
\begin{eqnarray}\label{eq:ExcLDA}
 E_{\mathrm{xc}}(\bi{n}) \approx E_{\mathrm{xc}}^{\mathrm{LDA}}(\bi{n}) := \sum_{l=1}^{L}e_{\mathrm{xc}}^{\mathrm{LDA}}(n_{l}) \qquad .
\end{eqnarray}
Because $E_{\mathrm{xc}}^{\mathrm{LDA}}(\bi{n})$ is the exact xc energy for exactly homogeneous densities in the thermodynamic limit, it represents a good approximation for relatively homogeneous densities on large lattices $L >> 1$.

\begin{figure}
\centering
\includegraphics[width=0.55\textwidth]{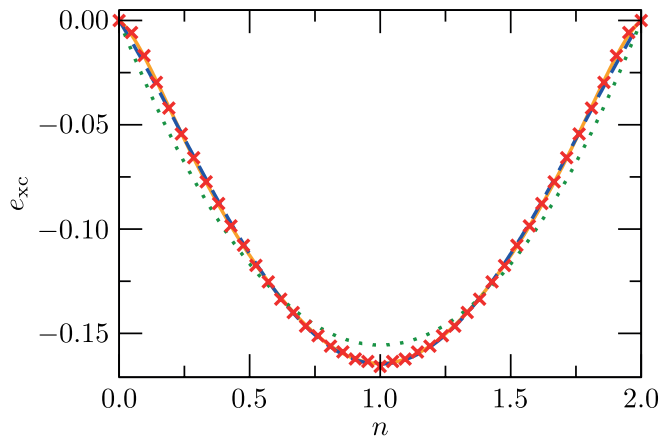}
\caption{\label{fig:1}
              Our exact LDA $e_{\mathrm{xc}}(n)$ (crosses) and polynomial interpolations $p_{d}(n)$ of degree $d=2$ (dotted), $4$ (dashed), and $8$ (solid).
             }
\end{figure}

Our feasibility study here assumes a relatively small lattice of size $L=21$ with hard-wall boundary conditions, such that finite size and boundary effects play a role.
We therefore derive our own LDA for this system and do not make use of the existing results in~\cite{Helbig}.
\Fref{fig:1} shows our $e_{\mathrm{xc}}(n)$ obtained from numerically exactly homogeneous densities computed by means of non-interacting (for $T^{\mathrm{s}}(n)$) and interacting (for $F_{\mathrm{HK}}(n)$) inversions on $L=21$ lattice sites for all possible total particle numbers $N=0,1,\ldots,42$.
To allow for an efficient evaluation of $E_{\mathrm{xc}}^{\mathrm{LDA}}$, we parametrize the function $e_{\mathrm{xc}}(n)$ using a finite number of parameters.
A simple way to achieve this is to assume a polynomial form $p_{d}(n)$ of certain degree $d$ and to fit our results using different values of $d$.
In the fit of each polynomial $p_{d}(n)$, we impose the physically reasonable constraint $p_{d}(0)=0=p_{d}(2)$, which trivially holds for the exact $E_{\mathrm{xc}}$, as can be seen in~\eref{eq:Exc}:
Obviously $e_{\mathrm{xc}}(0)=E_{\mathrm{xc}}(0, 0, \ldots, 0)/L=0$ because every term in~\eref{eq:Exc} vanishes independently for zero total particle number, and $e_{\mathrm{xc}}(2)=E_{\mathrm{xc}}(2, 2, \ldots, 2)/L=0$ because $F_{\mathrm{HK}}(2, 2, \ldots, 2)=E_{\mathrm{H}}(2, 2, \ldots, 2)$ and $T^{\mathrm{s}}(2, 2, \ldots, 2)=0$ due to impossible tunneling.
Apparently, our exact LDA $e_{\mathrm{xc}}$ is well approximated by polynomials of low degree $d$ since, on the scale of \fref{fig:1}, the $d=8$ fit seems indistinguishable from the $d=4$ fit.

The LDA resides on the lowest rung of ``Jacob's ladder''~\cite{Perdew5} and the most successful approximations of $E_{\mathrm{xc}}$ beyond the LDA were built on top of it~\cite{Perdew, Lee, Becke, Perdew2, Becke2, Perdew3, Zhao}.
Analogously, we will use the LDA computed above as our reference, and we will try to improve upon it with a more general ansatz for the functional.

\section{Our ansatz}
\label{sec:ouransatz}

Our approach for the construction of an improved xc energy approximation consists of two parts.
Firstly, it requires an efficient variational density functional ansatz, denoted by $G$, to approximate $E_{\mathrm{xc}}$.
Secondly, a set of $M$ external potentials has to be specified, which will be called \emph{training scenario}, such that the corresponding $M$ exact ground state densities $\bi{n}^{t}$, called \emph{training densities}, and exact values $E_{\mathrm{xc}}(\bi{n}^{t})$ are used to determine $G$ by minimizing a cost function
\begin{eqnarray}\label{eq:costFunction}
 \mathrm{d}(G) := \sum_{t=1}^{M}|E_{\mathrm{xc}}(\bi{n}^{t})-G(\bi{n}^{t})|^{2}
\end{eqnarray}
over the variational parameters of $G$.

We are interested in an ansatz that, firstly, includes the LDA and, secondly, allows for a systematic improvement over it by including non-local terms.
In this spirit, we propose a two-site ansatz of the following form:
\begin{eqnarray}\label{eq:ourAnsatz}
 G^{X}(\bi{n}) := \sum_{k=0}^{X} G^{k}(\bi{n})
\end{eqnarray}
with
\numparts\begin{eqnarray}
 G^{k=0}(\bi{n}) := \sum_{l=1}^{L} g^{0}(n_{l})\\
 G^{k>0}(\bi{n}) := \sum_{l=1}^{L-k} g^{k}(n_{l}, n_{l+k}) \qquad .
\end{eqnarray}\endnumparts
For $X=0$, we have $G(\bi{n})=G^{X=0}(\bi{n})=G^{k=0}(\bi{n})$, which is completely analogous to the LDA~\eref{eq:ExcLDA}.
And for $X>0$, the $k>0$ terms allow for a more general dependence on the density with two-site functions over a range limited by $X$.
In this way, increasing $X$ allows us to systematically include more non-local information and to go beyond the local LDA.

In order to have a practical functional, we want to write it in terms of a discrete set of variational parameters.
Thus we need to restrict the form of the functions $g^{k}$.
For simplicity we choose here a polynomial form for each term, as we did in the previous \sref{sec:exactLDA} for the reference LDA.
Additionally, in all following numerical experiments, we simply fix the degree of the polynomial to $d=4$.

When $G(\bi{n})$ is assumed to be a polynomial of the $n_{l}$, the variational parameters of $G$ are the polynomial coefficients.
Then the desired $\mathrm{argmin}_{G}\mathrm{d}(G)$, i.e.\ the argument of $\mathrm{d}$ that minimizes the cost function, results from the solution of linear equations $A\bi{c}=\bi{E}_{\mathrm{xc}}$ where the polynomial coefficients of $G$ are vectorized in $\bi{c}$, the exact values are vectorized in $\bi{E}_{\mathrm{xc}}$, and the elements in the matrix $A$ establish the correct connection to the cost function~\eref{eq:costFunction}: $\mathrm{d}(G)=\sum_{t=1}^{M}|E_{\mathrm{xc}}(\bi{n}^{t})-G(\bi{n}^{t})|^{2}=\sum_{t=1}^{M}|(\bi{E}_{\mathrm{xc}})_{t}-(A\bi{c})_{t}|^{2}$.

We want to emphasize that this approach does not need any computationally demanding interacting inversion.
Because the training densities $\bi{n}^{t}$ follow from the training potentials, i.e.\ from $M$ different choices of $\bi{v}^{\mathrm{ext}}$ in~\eref{eq:funcFHK}, we know $F_{\mathrm{HK}}(\bi{n}^{t})$.
While the calculation of $E_{\mathrm{H}}(\bi{n}^{t})$ is trivial, $T^{\mathrm{s}}(\bi{n}^{t})$ is computed via efficient non-interacting inversion.
Thus, all further ingredients for $E_{\mathrm{xc}}(\bi{n}^{t})$ of~\eref{eq:Exc} are then efficiently computable.

\begin{figure}
\centering
\includegraphics[width=0.55\textwidth]{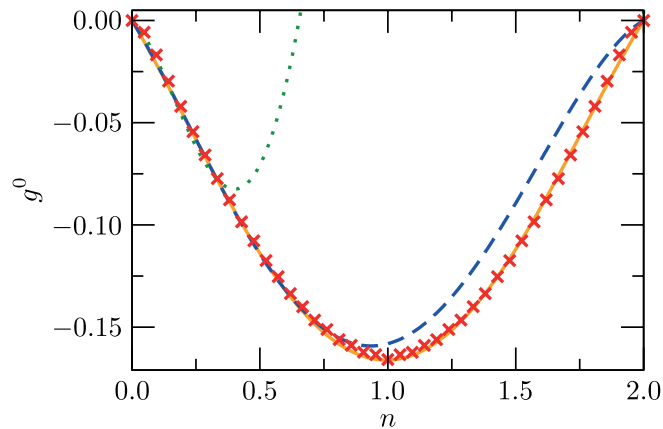}
\caption{\label{fig:2}
              Local terms $g^{0}(n)$ from the ``homogeneous'' training scenario with $M=N=6$ (dotted), $12$ (dashed), and $30$ (solid), compared to the exact LDA (crosses).
              Here, $g^{0}$ is a polynomial of degree $d=4$.
             }
\end{figure}

The first step in our approach is to consider the ansatz $G=G^{0}$, with $g^{0}(n_{l}) := \sum_{s=0}^{d}c_{s}^{0}n_{l}^{s}$ a polynomial in $n_{l}$ of degree $d$ with coefficients $c_{s}^{0}$ (and $d=4$ in the following).
The simplest possible, ``homogeneous'', training scenario amounts to setting $v_{l}^{\mathrm{ext}}=0$, in which case we have one training density for each possible total particle number $N=1,2,\ldots,2L$, i.e.\ at most $M=42$ for $L=21$.
\Fref{fig:2} demonstrates how training our local term $g^{0}$ with such ground states reproduces the exact LDA.
Remarkably, a very good match between our $g^{0}$ and the exact LDA is achieved already with $M=30$ ground state computations.
This has to be compared to the several thousands of ground state computations that were required for \fref{fig:1} due to the interacting inversion iteration.

\begin{figure}
\centering
\includegraphics[width=0.506646\textwidth]{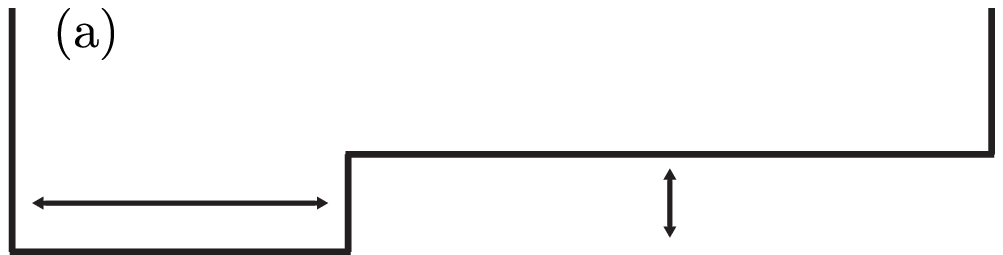}\\[5mm]
\includegraphics[width=0.506646\textwidth]{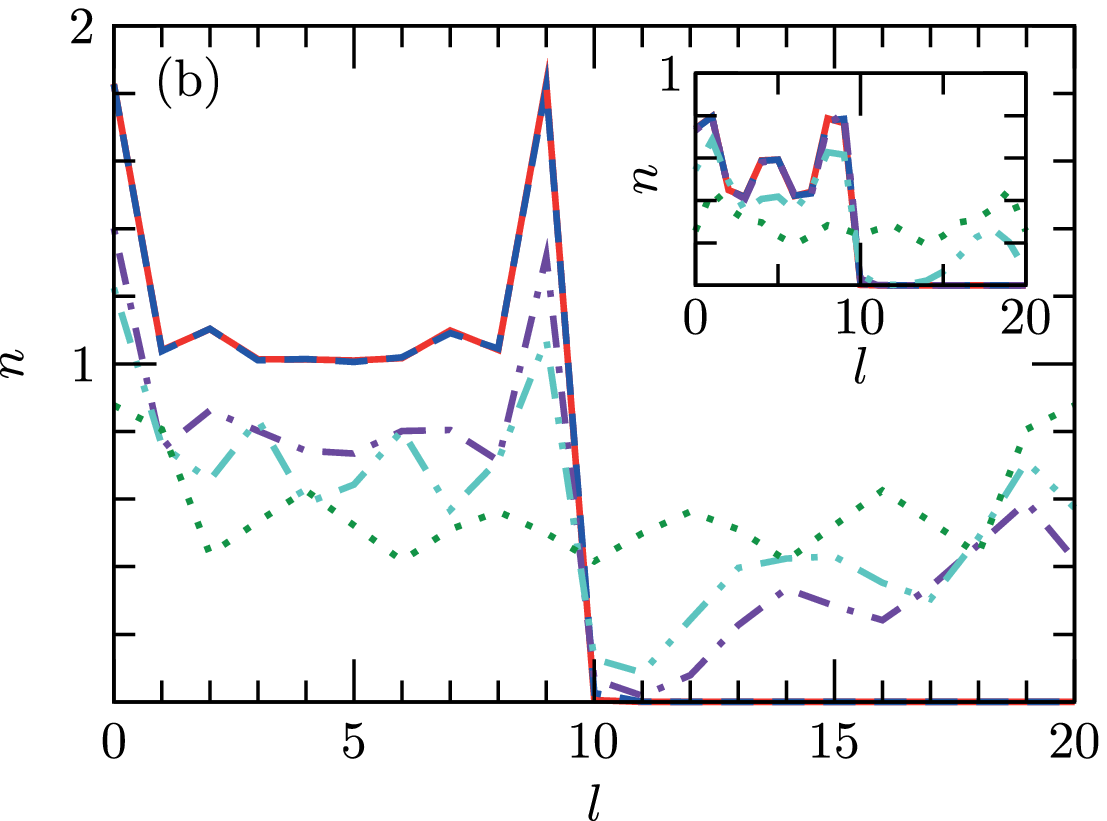}
\caption{\label{fig:3}
              (a) ``Step'' training scenario: external potentials $v_{l}^{\mathrm{ext}}$ are characterized by a step of certain height at a certain position.
              (b) Ground state density $n_{l}$ for $N=12$ (main) and $6$ (inset), for a step at position $l=10$ of height $h=0$ (dotted), $0.3$ (dash-double dotted), $0.5$ (dash-dotted), $1.0$ (dashed), and $2.0$ (solid).
             }
\end{figure}

We can understand that the ``homogeneous'' training scenario leads to the exact LDA because this scenario contains relatively homogeneous training densities and because the exact LDA is constructed from exactly homogeneous densities.
Now we want to consider more inhomogeneous densities.
For that purpose we propose the simplest possible extension of the ``homogeneous'' training scenario: the ``step'' training scenario shown in \fref{fig:3} (a).
This scenario contains the simplest external potentials that give rise to inhomogeneous ground state densities, see \fref{fig:3} (b) for some example densities.
The ``step'' training scenario allows us to generate much larger training sets since we define it by free choice of: a) the step position (from $l=2, 3, 4, \ldots, 21$), b) the step height (from $h=0, 0.1, 0.2, \ldots, 2.0$), c) the step orientation (\emph{left} or \emph{right}), and d) the total particle number (from $N=1, 2, 3, \ldots, 30$).
We do not include total particle numbers $N$ larger than $30$ in this training set because we want it to contain sufficiently inhomogeneous densities that become more inhomogeneous when the step height increases; clearly, for large total particle numbers such as more than $30$ fermions on $21$ lattice sites, an increasing step height quickly creates large homogeneous regions of maximum filling in the density, i.e.\ having $2$ fermions per lattice site.

We have investigated two different ways of converging $G$ with this ``step'' training scenario.
In the first way, we pick $M$ ground states randomly and study the convergence of $G$ as a function of $M$.
In the second way, we fix the total particle numbers considered to $N=1,2,\ldots,12$, take all possible step positions and orientations, and increase $M$ systematically together with the step height.
In both schemes, convergence is quantified by comparison of the solution for $M$ with the solution for the largest considered $M_{\mathrm{max}}$, which we fixed to $12800$ for the random and to $9612$ for the systematic densities.
We can then look at the quantity
\begin{eqnarray}
 \epsilon(M) := \langle|C_{i}(M)-C_{i}(M_{\mathrm{max}})|/|C_{i}(M_{\mathrm{max}})|\rangle
\end{eqnarray}
where $C_{i}(M)$ denotes the $i$th parameter of $G$ after training with $M$ densities and $\langle \ldots \rangle := 1/P\sum_{i=1}^{P}\ldots$ denotes taking the mean value over all $P$ possible values of $i$.

\begin{figure}
\centering
\includegraphics[width=0.55\textwidth]{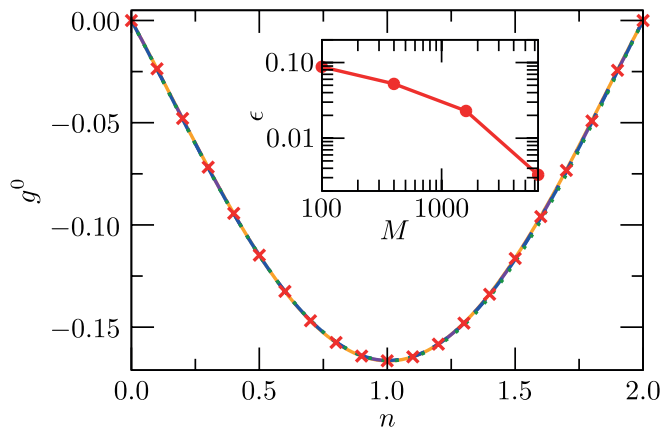}
\caption{\label{fig:4}
              Local terms $g^{0}(n)$ from the ``step'' training scenario with $M=100$ (dotted), $400$ (dash-dotted), $1600$ (dashed), $6400$ (solid), and $12800$ (crosses).
              Inset: Mean relative difference $\epsilon(M)$ between the coefficients of $g^{0}$ after training with $M$ densities and the coefficients of $g^{0}$ after training with $M_{\mathrm{max}}=12800$ densities.
              Here, $g^{0}$ is a polynomial of degree $d=4$.
             }
\end{figure}

\Fref{fig:4} shows our results for random densities.
Interestingly, this training scenario gives rise to local terms $g^{0}$ that are very similar to the exact LDA, although many training densities of this scenario are very inhomogeneous.
Furthermore, we can read off from the inset of \fref{fig:4} that convergence occurs rapidly.

To go beyond the LDA, we now include the longer-range two-site terms with $k>0$ in~\eref{eq:ourAnsatz} using a general polynomial ansatz:
\begin{eqnarray}
 g^{k}(n_{l}, n_{l+k}) := \sum_{s_{0},s_{k}=0}^{d} c_{s_{0},s_{k}}^{k} n_{l}^{s_{0}} n_{l+k}^{s_{k}} \quad ,
\end{eqnarray}
i.e.\ these terms are general degree $d$ polynomials of the density values on $2$ lattice sites separated by distance $k$ (and, again, we simply fix $d=4$ in the following).
While, as discussed above, for $X=0$, our ansatz $G^{0}(\bi{n})=\sum_{l=1}^{L}g^{0}(n_{l})$ is completely analogous to the LDA of~\eref{eq:ExcLDA}, for $X>0$, it contains additional non-local terms, such that, by systematically increasing $X$ in our ansatz $G^{X}(\bi{n})$, we can systematically increase its non-locality beyond the local LDA-like term.

We enforce in our desired solution $G^{X}$ that the terms $g^{k}$ for increasing $k$ are obtained one after another such that each additional non-local term (i.e.\ corresponding to the next larger value of $k$) is a correction to the previous solution.
This means that, for given $X$, we first minimize~\eref{eq:costFunction} only via the parameters $c^{0}$ which gives $G^{0}$.
Then we minimize~\eref{eq:costFunction} for the remainder $E_{\mathrm{xc}}^{1}(\bi{n}^{t}):=E_{\mathrm{xc}}(\bi{n}^{t})-G^{0}(\bi{n}^{t})$ only via the parameters $c^{1}$ which together with the previous solution $c^{0}$ gives $G^{1}$.
Then we minimize~\eref{eq:costFunction} for the remainder $E_{\mathrm{xc}}^{2}(\bi{n}^{t}):=E_{\mathrm{xc}}(\bi{n}^{t})-G^{1}(\bi{n}^{t})$ only via the parameters $c^{2}$ which together with the previous solutions $c^{0}$ and $c^{1}$ gives $G^{1}$.
We continue the scheme until we have reached $c^{X}$ and thus $G^{X}$.
This procedure ensures that each longer-range term is built on top of all previous shorter-ranged ones, in the same way as the functionals on higher rungs of ``Jacob's ladder'' are more non-local and are built on top of the more local functionals on the lower rungs~\cite{Perdew5}.

In order to analyze the performance of the ansatz, we adopt now a different strategy.
We will now always fit our ansatz $G^{X}$ with the ``step'' training scenario of \fref{fig:3} and then we will apply it to completely different target densities.
For the latter, we choose ground states of the $H_{2}$ dissociation problem, i.e.\ Hamiltonian~\eref{eq:ham} with total particle number $N=2$ and external potential
\numparts\begin{eqnarray}
 \hat{V}(R) := \sum_{l=1}^{L}v_{l}^{\mathrm{ext}}(R)\hat{n}_{l}+\frac{1}{\sqrt{R^{2}+1}} \\
 v_{l}^{\mathrm{ext}}(R) := -\frac{1}{\sqrt{(l-(l_{0}-R/2))^{2}+1}} - \frac{1}{\sqrt{(l-(l_{0}+R/2))^{2}+1}}
\end{eqnarray}\endnumparts
where $R$ denotes the separation between the two $H$ atoms placed in the middle of the lattice such that we set $l_{0}=11$ for $L=21$.
Because this problem represents a realistic physical application that is significantly different from our training scenario, we consider it to be a very good benchmark for our approach.

From now on, the local terms $g^{0}(n)$ in our two-site polynomial ansatz~\eref{eq:ourAnsatz} are fixed to be the exact LDA from \fref{fig:1}.
And the non-local terms $g^{k>0}(n_{1}, n_{2})$ are enforced to fulfill $g^{k>0}(0,0)=0=g^{k>0}(2,2)$ as well as $g^{k>0}(n_{1}, n_{2})=g^{k>0}(n_{2}, n_{1})$.
These properties are physically reasonable and they reduce the number of variational parameters, which turned out to be beneficial for the convergence of our fit.
In particular, this helped us to avoid the effect known as overfitting.
We distinguish two different versions of our ansatz, namely \emph{constrained} and \emph{unconstrained}.
In our constrained two-site polynomial ansatz we determine $g^{k>0}(n_{1}, n_{2})$ under the constraint $g^{k>0}(n, n)=0$: then $G^{X}(\bi{n})$ is exact for exactly homogeneous densities $\bi{n}=(n, n, \ldots, n)^{\mathrm{T}}$.
In our unconstrained two-site polynomial ansatz we do not impose this constraint on the polynomial coefficients: the unconstrained ansatz has thus more variational parameters than the constrained ansatz.

\begin{figure}
\centering
\includegraphics[width=0.55\textwidth]{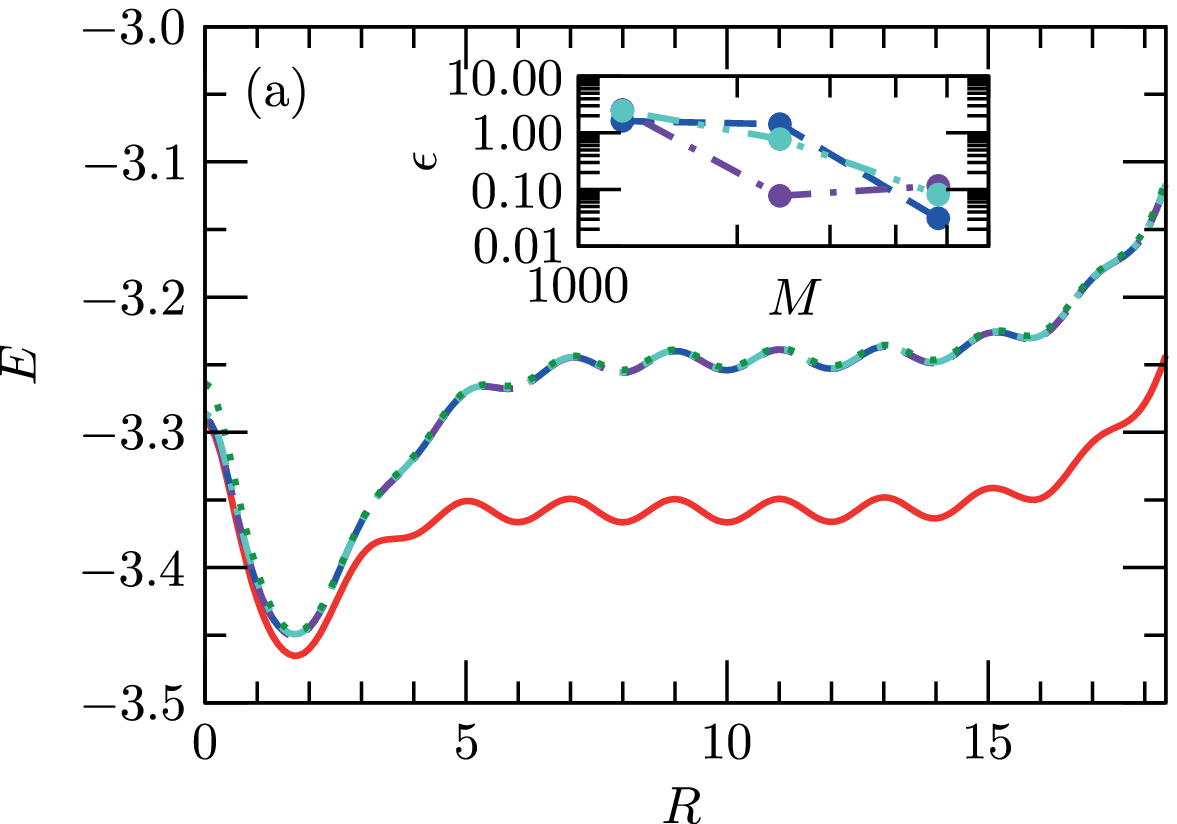}
\includegraphics[width=0.55\textwidth]{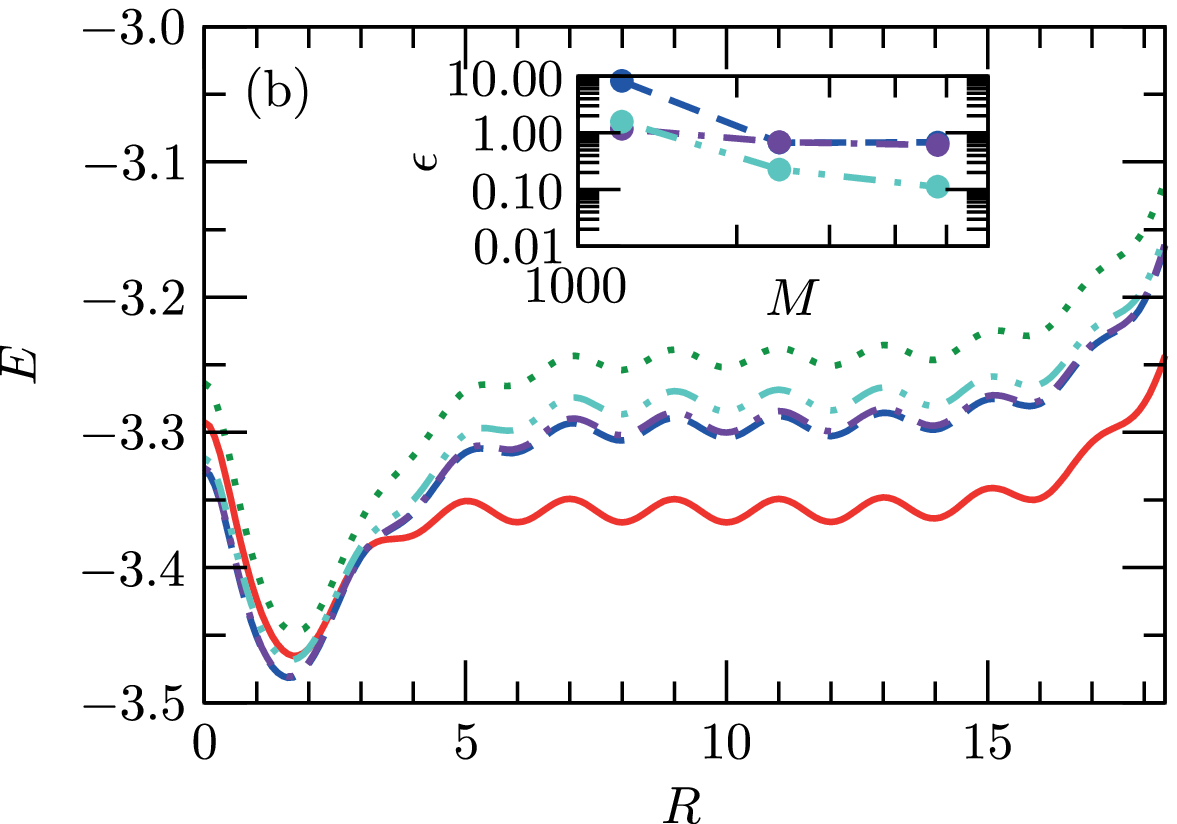}
\caption{\label{fig:5}
              $H_{2}$ dissociation energy $E(R)$ as a function of the separation $R$: from exact LDA (dotted) and from our constrained (a) and unconstrained (b) ansatz with $X=1$ (dash-double dotted), $2$ (dash-dotted), $5$ (dashed),
              compared to the exact solution (solid).
              Insets: Mean relative difference $\epsilon(M)$ between the coefficients of $g^{k>0}$ for $k=1$ (dash-double dotted), $2$ (dash-dotted), and $5$ (dashed) after training with $M$ systematic densities and the corresponding coefficients
              of $g^{k>0}$ after training with $M_{\mathrm{max}}=9612$ systematic densities - the insets show our results for $M=1212$, $2412$, and $4812$.
              Here, our ansaetze for $G$ are polynomials of degree $d=4$.
            }
\end{figure}

\Fref{fig:5} shows our results after convergence with $M$ systematic densities.
As we can see in (a), the constrained ansatz leads to a visible improvement over LDA close to $R=0$, but not for larger $R$.
A convergence of the non-local terms can be concluded from the inset.
In (b), the unconstrained ansatz leads to an improvement over LDA for almost all values of $R$, but it produces too low energy values close to $R=0$.
The inset of (b) demonstrates that convergence of the non-local terms occurs, however, for $k>1$ this convergence is slower than in (a).
With increasing $X$, our ansatz systematically improves the LDA result at specific values of $R$:
around $R=0$ when the constrained version is used, and at larger $R$ when the unconstrained version is used.
Both versions of our ansatz show a systematic improvement over LDA with increasing $X$ when the mean energy for all values of $R$ is considered.
In fact, such a mean value is the correct figure of merit because the cost function~\eref{eq:costFunction}, minimized for ``step'' training densities by our ansatz, is also a mean value of many xc energies.

\begin{figure}
\centering
\includegraphics[width=0.55\textwidth]{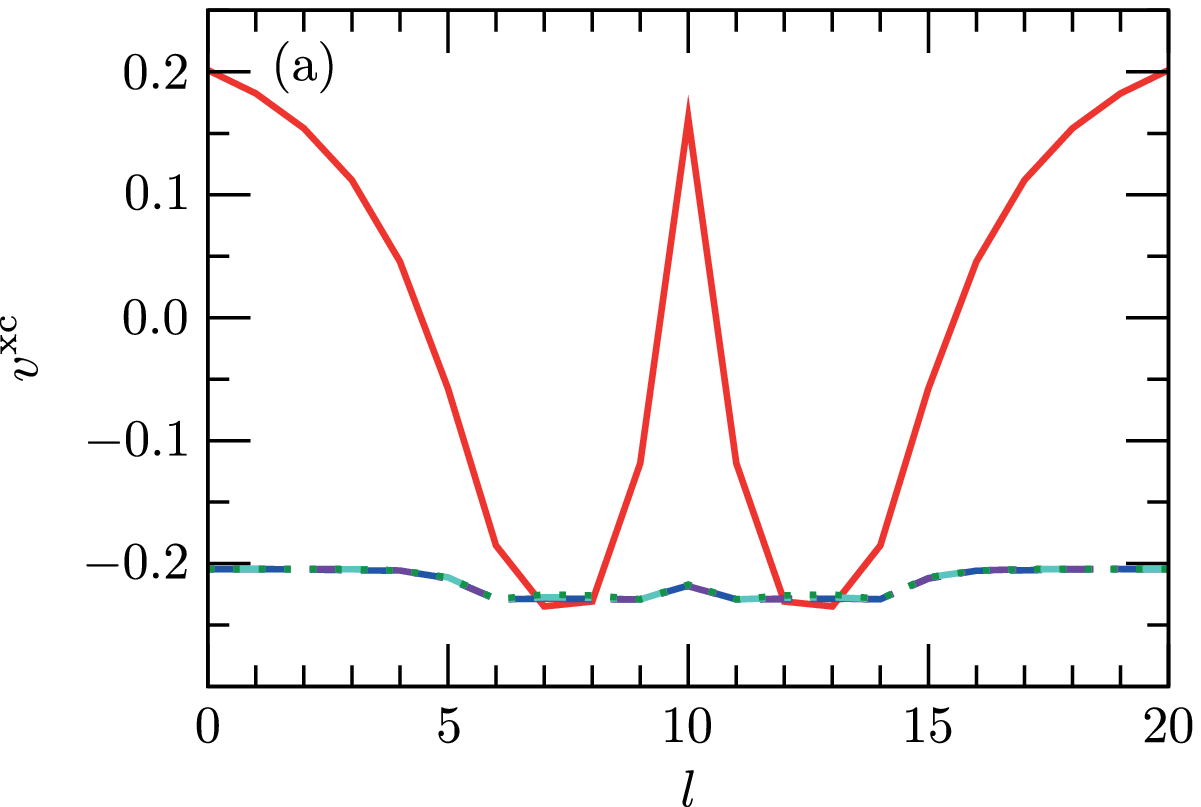}
\includegraphics[width=0.55\textwidth]{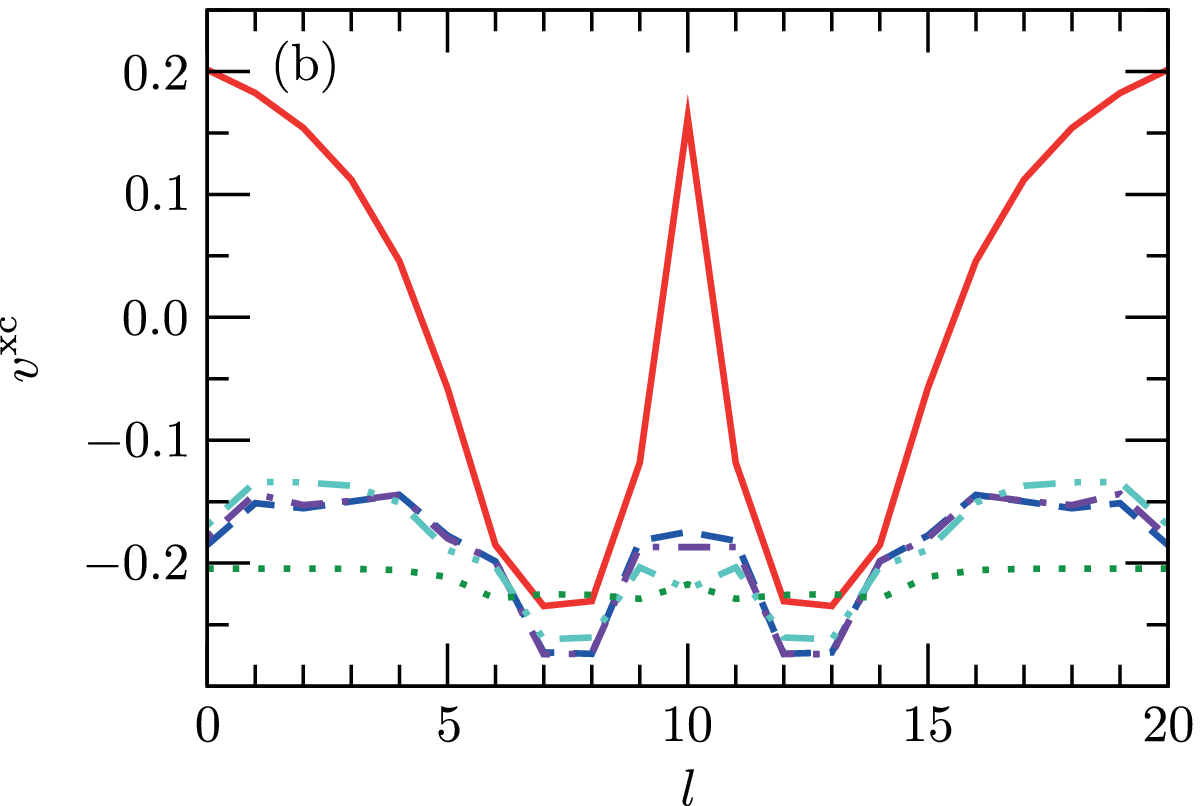}
\caption{\label{fig:6}
              Exchange-correlation potential $v_{l}^{\mathrm{xc}}$ for the exact ground state density of the $H_{2}$ dissociation curve at $R=5$: from exact LDA (dotted) and from our constrained (a) and unconstrained (b) ansatz with
              $X=1$ (dash-double dotted), $2$ (dash-dotted), $5$ (dashed), compared to the exact solution (solid).
              Our ansaetze for $G$ here are the same as the ones in \fref{fig:5}.
             }
\end{figure}

Clearly, we would like to use $G^{X}(\bi{n})$ to compute densities self-consistently via the KS cycle.
A first step in this direction is the calculation of the xc potential $v^{\mathrm{xc}}_{l}(\bi{n}):=-\partial E_{\mathrm{xc}} / \partial n_{l} |_{\bi{n}}$ for the exact ground state density $\bi{n}$.
In the $H_{2}$ dissociation problem, the xc potential for larger values of $R$ is particularly interesting, since its exact form exhibits a characteristic peak that cannot be reproduced by LDA alone~\cite{Fuchs}.
\Fref{fig:6} shows our results for $R=5$.
While our constrained ansatz leads to a potential that basically coincides with the one from LDA, our unconstrained ansatz leads to a small systematic improvement with increasing $X$.

\section{Conclusions}
\label{sec:conclusions}

We have analyzed the feasibility of constructing semi-empirical approximations for the xc density functional in the context of a long-range interacting many-electron system on a one-dimensional lattice.
Using numerically exact ground states from MPS simulations, we proposed to fit an ansatz that includes an LDA-like part plus additional terms of increasing non-locality, by means of reasonably chosen training densities.
We observed that our ansatz converges systematically within the training scenario.
Additionally, when applied to completely different target densities, namely of the $H_{2}$ dissociation problem, our fitted ansatz improved upon the LDA systematically.
This systematic improvement was demonstrated for the ground state energies of the $H_{2}$ problem and for a xc potential corresponding to a stretched $H_{2}$ molecule.

In this work, we have tested the effect of a systematic inclusion of non-local ingredients in the functional using very simple ansaetze for the non-local terms, namely only two-site dependences, and for the functional form, namely only polynomials, and we considered only one training scenario.
Our results show that by systematically including non-local terms, the approximation can without doubt be improved beyond LDA.
Our quantitative results are nevertheless limited to the specific form of the ansatz and training set used.
For instance, the fact that the dissociation curve does not significantly improve by including terms of longer range after some point seems to indicate that other non-local contributions may be more relevant.
Likewise, considering functional forms for each term that go beyond a polynomial may improve the power of the ansatz.
We have already run initial tests to study the effect of terms that depend on three variable densities, but we observe no clear convergence with as many as $\approx 20000$ densities from the ``step'' training scenario.
This clearly indicates the necessity for a different training scenario and possibly additional physical constraints that effectively reduce the number of variational parameters.
The question itself of how to choose the training densities optimally is, in general, a very important one that should be further explored, as a better training scenario would always further improve our results.
All in all, although such improved ansaetze and training scenarios must definitely achieve better results than the ones reported here, a careful analysis is beyond the scope of this proof-of-principle work.

It would be very interesting to combine our procedure with concepts from recent works on machine learning of density functionals~\cite{Snyder, Snyder2, Snyder3, Snyder4, Vu, Li}.
On the one hand, these works typically required less training densities than our approach.
On the other hand, our work constructs systematic corrections to a standard approximation, namely the LDA, that can be applied in general, i.e.\ to other types of systems beyond those that were originally used for the fit.
Thus, a combination of the good aspects of our procedure with the good aspects of the previous machine learning concepts could be the ultimate solution to some of the problems that both approaches currently have independently from each other.

In this article, we focused exclusively on approximations of the density functional for the xc energy.
However, in principle, any ground state observable can be written as a functional of the ground state density~\cite{Hohenberg}.
Our proposed scheme allows in principle to also construct systematically density functionals for observables other than the ground state energy.
This fact might now be useful for ultracold atoms in optical lattices since the densities of these systems have become experimentally accessible with remarkable resolution~\cite{Bakr, Sherson, Cheuk, Parsons, Miranda, Haller, Edge, Omran}.
Measurements of a certain observable which are difficult to perform with the current techniques might be easier to carry out now when a density functional would be provided for that observable.
Particularly interesting measurements regard two-dimensional systems with special observables, such as e.g.\ special order parameters.
The corresponding density functionals could be constructed with the help of projected entangled pair states (PEPS)~\cite{Verstraete2}.
Because experiments are restricted to finite system sizes, PEPS algorithms are perfectly suited for the ground state simulation of such finite two-dimensional systems, for which they can produce accurate numerical results~\cite{Murg, Murg2, Lubasch, Hosseinkhani, Lubasch2, Lubasch3}.

\ack

M L is very grateful to Neepa T Maitra and Garnet K-L Chan for discussions.
He acknowledges funding by the EU through SIQS grant (FP7 600645) and the DFG (NIM cluster of excellence).
He also thanks the Pedro Pascual Benasque Center for Science (CCBPP), where he carried out part of this work.
A R acknowledges financial support from the European Research Council (ERC-2010-AdG-267374), Spanish grant (FIS2013-46159-C3-1-P), Grupos Consolidados (IT578-13).


\section*{References}

\begin{thebibliography}{99}

\bibitem{Hohenberg} Hohenberg P and Kohn W 1964 Inhomogeneous Electron Gas \PR {\bf 136} B864--71

\bibitem{Kohn} Kohn W and Sham L J 1965 Self-Consistent Equations Including Exchange and Correlation Effects \PR {\bf 140} A1133--38

\bibitem{Dreizler} Dreizler R M and Gross E K U 1990 {\it Density Functional Theory: An Approach to the Quantum Many-Body Problem} (Berlin, Heidelberg: Springer-Verlag)

\bibitem{Kohn2} Kohn W 1999 Nobel Lecture: Electronic structure of matter -- wave functions and density functionals \RMP {\bf 71} 1253-66

\bibitem{Jones} Jones R O 2015 Density functional theory: Its origins, rise to prominence, and future \RMP {\bf 87} 897--923

\bibitem{Perdew} Perdew J P and Zunger A 1981 Self-interaction correction to density-functional approximations for many-electron systems \PR B {\bf 23} 5048--79

\bibitem{Lee} Lee C, Yang W and Parr R G 1988 Development of the Colle-Salvetti correlation-energy formula into a functional of the electron density \PR B {\bf 37} 785--9

\bibitem{Becke} Becke A D 1988 Density-functional exchange-energy approximation with correct asymptotic behavior \PR A {\bf 38} 3098--100

\bibitem{Perdew2} Perdew J P and Wang Y 1992 Accurate and simple analytic representation of the electron-gas correlation energy \PR B {\bf 45} 13244--13249

\bibitem{Becke2} Becke A D 1993 Density-functional thermochemistry. III. The role of exact exchange \JCP {\bf 98} 5648--52

\bibitem{Perdew3} Perdew J P, Burke K and Ernzerhof M 1996 Generalized Gradient Approximation Made Simple \PRL {\bf 77} 3865--8

\bibitem{Zhao} Zhao Y and Truhlar D G 2008 The M06 suite of density functionals for main group thermochemistry, thermochemical kinetics, noncovalent interactions, excited states, and transition elements: two new functionals and systematic testing of four M06-class functionals and 12 other functionals {\it Theor.\ Chem.\ Accounts} {\bf 120} 215--41

\bibitem{Schuch} Schuch N and Verstraete F 2009 Computational complexity of interacting electrons and fundamental limitations of density functional theory {\it Nature Physics} {\bf 5} 732--5

\bibitem{Peverati} Peverati R and Truhlar D G 2013 The quest for a universal density functional: The accuracy of density functionals across a broad spectrum of databases in chemistry and physics arXiv:1212.0944v4 [physics.chem-ph]

\bibitem{Perdew4} Perdew J P and Yue W 1986 Accurate and simple density functional for the electronic exchange energy: Generalized gradient approximation \PR B {\bf 33} 8800--2(R)

\bibitem{Kurth} Kurth S and Perdew J P 2000 Role of the Exchange-Correlation Energy: Nature's Glue {\it Int.\ J.\ Quantum Chem.\ } {\bf 77} 814--8

\bibitem{Perdew5} Perdew J P, Ruzsinszky A, Tao J, Staroverov V N, Scuseria G E and Csonka G I 2005 Prescription for the design and selection of density functional approximations: More constraint satisfaction with fewer fits \JCP {\bf 123} 062201-1--062201-9

\bibitem{Verstraete} Verstraete F, Murg V and Cirac J I 2008 Matrix product states, projected entangled pair states, and variational renormalization group methods for quantum spin systems {\it Adv.\ Phys.\ } {\bf 57} 143--224

\bibitem{Orus} Or\'{u}s R 2014 A practical introduction to tensor networks: Matrix product states and projected entangled pair states {\it Ann.\ Phys.\ } {\bf 349} 117--158

\bibitem{Giuliani} Giuliani G F and Vignale G 2005 {\it Quantum Theory of the Electron Liquid} (Cambridge: Cambridge University Press)

\bibitem{White} White S R 1992 Density Matrix Formulation for Quantum Renormalization Groups \PRL {\bf 69} 2863--6

\bibitem{Schollwoeck} Schollw\"{o}ck U 2011 The density-matrix renormalization group in the age of matrix product states {\it Ann.\ Phys.\ } {\bf 326} 96--192

\bibitem{Stoudenmire} Stoudenmire E M, Wagner L O, White S R and Burke K 2012 One-Dimensional Continuum Electronic Structure with the Density-Matrix Renormalization Group and Its Implications for Density-Functional Theory \PRL {\bf 109} 056402-1--056402-5

\bibitem{Wagner} Wagner L O, Stoudenmire E M, Burke K and White S R 2012 Reference electronic structure calculations in one dimension {\it Phys.\ Chem.\ Chem.\ Phys.\ } {\bf 14} 8581--90

\bibitem{Peirs} Peirs K, Van Neck D and Waroquier M 2003 Algorithm to derive exact exchange-correlation potentials from correlated densities in atoms \PR A {\bf 67} 012505-1--012505-12

\bibitem{Thiele} Thiele M, Gross E K U and K\"{u}mmel S 2008 Adiabatic Approximation in Nonperturbative Time-Dependent Density-Functional Theory \PRL {\bf 100} 153004-1--153004-4

\bibitem{Stoudenmire2} Stoudenmire E M, Wagner L O, White S R and Burke K 2011 Exact density functional theory with the density matrix renormalization group arXiv:1107.2394v1 [cond-mat.str-el]

\bibitem{Wagner2} Wagner L O, Stoudenmire E M, Burke K and White S R 2013 Guaranteed Convergence of the Kohn-Sham Equations \PRL {\bf 111} 093003-1--093003-5

\bibitem{Helbig} Helbig N, Fuks J I, Casula M, Verstraete M J, Marques M A L, Tokatly I V and Rubio A 2011 Density functional theory beyond the linear regime: Validating an adiabatic local density approximation \PR A {\bf 83} 032503-1--032503-5

\bibitem{Fuchs} Fuchs M, Niquet Y-M, Gonze X and Burke K 2005 Describing static correlation in bond dissociation by Kohn--Sham density functional theory \JCP {\bf 122} 094116-1--094116-13

\bibitem{Snyder} Snyder J C, Rupp M, Hansen K, M\"{u}ller K-R and Burke K 2012 Finding Density Functionals with Machine Learning \PRL {\bf 108} 253002-1--253002-5

\bibitem{Snyder2} Snyder J C, Mika S, Burke K and M\"{u}ller K-R 2013 Kernels, Pre-Images and Optimization {\it Chapter in Empirical Inference - Festschrift in Honor of Vladimir N Vapnik}

\bibitem{Snyder3} Snyder J C, Rupp M, Hansen K, Blooston L, M\"{u}ller K-R and Burke K 2013 Orbital-free bond breaking via machine learning \JCP {\bf 139} 224104-1--224104-10

\bibitem{Snyder4} Snyder J C, Rupp M, M\"{u}ller K-R and Burke K 2015 Nonlinear Gradient Denoising: Finding Accurate Extrema from Inaccurate Functional Derivatives {\it Int.\ J.\ Quantum Chem.\ } {\bf 2015} 1102--14

\bibitem{Vu} Vu K, Snyder J C, Li L, Rupp M, Chen B F, Khelif T, M\"{u}ller K-R and Burke K 2015 Understanding Kernel Ridge Regression: Common Behaviors from Simple Functions to Density Functionals {\it Int.\ J.\ Quantum Chem.\ } {\bf 2015} 1115--28

\bibitem{Li} Li L, Snyder J C, Pelaschier I M, Huang J, Niranjan U-N, Duncan P, Rupp M, M\"{u}ller K-R and Burke K 2015 Understanding Machine-Learned Density Functionals {\it Int.\ J.\ Quantum Chem.\ } {\bf 2016} 819--33

\bibitem{Bakr} Bakr W S, Gillen J I, Peng A, F\"{o}lling S and Greiner M 2009 A quantum gas microscope for detecting single atoms in a Hubbard-regime optical lattice {\it Nature} {\bf 462} 74--7

\bibitem{Sherson} Sherson J F, Weitenberg Ch, Endres M, Cheneau M, Bloch I and Kuhr S 2010 Single-atom-resolved fluorescence imaging of an atomic Mott insulator {\it Nature} {\bf 467} 68--72

\bibitem{Cheuk} Cheuk L W, Nichols M A, Okan M, Gersdorf T, Ramasesh V V, Bakr W S, Lompe T and Zwierlein M W 2015 Quantum-Gas Microscope for Fermionic Atoms \PRL {\bf 114} 193001-1--193001-5

\bibitem{Parsons} Parsons M F, Huber F, Mazurenko A, Chiu Ch S, Setiawan W, Wooley-Brown K, Blatt S and Greiner M 2015 Site-Resolved Imaging of Fermionic $^{6}$Li in an Optical Lattice \PRL {\bf 114} 213002-1--213002-5

\bibitem{Miranda} Miranda M, Inoue R, Okuyama Y, Nakamoto A and Kozuma M 2015 Site-resolved imaging of ytterbium atoms in a two-dimensional optical lattice \PR A {\bf 91} 063414-1--063414-6

\bibitem{Haller} Haller E, Hudson J, Kelly A, Cotta D A, Peaudecerf B, Bruce G D and Kuhr S 2015 Single-atom imaging of fermions in a quantum-gas microscope {\it Nature Physics} {\bf 11} 738--42

\bibitem{Edge} Edge G J A, Anderson R, Jervis D, McKay D C, Day R, Trotzky S and Thywissen J H 2015 Imaging and addressing of individual fermionic atoms in an optical lattice \PR A {\bf 92} 063406-1--063406-6

\bibitem{Omran} Omran A, Boll M, Hilker T A, Kleinlein K, Salomon G, Bloch I and Gross Ch 2015 Microscopic Observation of Pauli Blocking in Degenerate Fermionic Lattice Gases \PRL {\bf 115} 263001-1--263001-5

\bibitem{Verstraete2} Verstraete F and Cirac J I 2004 Renormalization Algorithms for Quantum Many-Body Systems in Two and Higher Dimensions arXiv:cond-mat/0407066v1 [cond-mat.str-el]

\bibitem{Murg} Murg V, Verstraete F and Cirac J I 2007 Variational study of hard-core bosons in a two-dimensional optical lattice using projected entangled pair states \PR A {\bf 75} 033605-1--033605-8

\bibitem{Murg2} Murg V, Verstraete F and Cirac J I 2009 Exploring frustrated spin systems using projected entangled pair states \PR B {\bf 79} 195119-1--195119-7

\bibitem{Lubasch} Lubasch M, Murg V, Schneider U, Cirac J I and Ba\~{n}uls M-C 2011 Adiabatic Preparation of a Heisenberg Antiferromagnet Using an Optical Superlattice \PRL {\bf 107} 165301-1--165301-5

\bibitem{Hosseinkhani} Hosseinkhani A, Dezfouli B G, Ghasemipour F, Rezakhani A T and Saberi H 2014 Uncontrolled disorder effects in fabricating photonic quantum simulators on a kagome geometry: A projected-entangled-pair-state versus exact-diagonalization analysis \PR A {\bf 89} 062324-1--062324-7

\bibitem{Lubasch2} Lubasch M, Cirac J I and Ba\~{n}uls M-C 2014 Unifying projected entangled pair state contractions \NJP {\bf 16} 033014

\bibitem{Lubasch3} Lubasch M, Cirac J I and Ba\~{n}uls M-C 2014 Algorithms for finite projected entangled pair states \PR B {\bf 90} 064425-1--064425-16

\end{thebibliography}
\end{document}